\documentclass[conference,a4paper,10pt]{IEEEtran}
\IEEEoverridecommandlockouts
\usepackage{cite}
\usepackage{amsmath,amssymb,amsfonts}
\usepackage{algorithmic}
\usepackage{graphicx}
\graphicspath{{./figures/}}
\usepackage{textcomp}
\usepackage{xcolor}
\usepackage{pbalance}
\def\BibTeX{{\rm B\kern-.05em{\sc i\kern-.025em b}\kern-.08em
    T\kern-.1667em\lower.7ex\hbox{E}\kern-.125emX}}
\let\oldtt\texttt
\renewcommand{\texttt}[1]{\small\oldtt{#1}}

\renewcommand{\verb}[1]{\small\oldverb#1}

\newlength{\footnoterulewidth} 
\newlength{\footnoteruleheight} 
\renewcommand{\footnoterule}{   
    \kern -3pt   \hrule width \footnoterulewidth height \footnoteruleheight   \kern \dimexpr 3pt - \footnoteruleheight \relax 
}

\begin{document}

\title{A System for Differentiation of Schizophrenia and Bipolar Disorder based on rsfMRI}

\author{
    \IEEEauthorblockN{
          Daniela Janeva, Stefan Krsteski, Matea Tashkovska, Nikola Jovanovski, Tomislav Kartalov, Dimitar Taskovski,\\Zoran Ivanovski, and Branislav Gerazov\\ 
    }

    \IEEEauthorblockA{
        \\
         Faculty of Electrical Engineering and Information Technologies, \\
         Ss. Cyril and Methodius University, Skopje, Macedonia 
        \\   
        \texttt{danielajaneva@hotmail.com}
    }
}

\maketitle

\begin{abstract}
Schizophrenia and bipolar disorder are debilitating psychiatric illnesses that can be challenging to diagnose accurately. 
The similarities between the diseases make it difficult to differentiate between them using traditional diagnostic tools. 
Recently, resting-state functional magnetic resonance imaging (rsfMRI) has emerged as a promising tool for the diagnosis of psychiatric disorders. 
This paper presents several methods for differentiating schizophrenia and bipolar disorder based on features extracted from rsfMRI data. 
The system that achieved the best results, uses 1D Convolutional Neural Networks to analyze patterns of Intrinsic Connectivity time courses obtained from rsfMRI and potentially identify biomarkers that distinguish between the two disorders. 
We evaluate the system's performance on a large dataset of patients with schizophrenia and bipolar disorder and demonstrate that the system achieves a 0.7078 Area Under Curve (AUC) score in differentiating patients with these disorders. 
Our results suggest that rsfMRI-based classification systems have great potential for improving the accuracy of psychiatric diagnoses and may ultimately lead to more effective treatments for patients with this disorder.
\end{abstract}


\section{Introduction}
Schizophrenia and bipolar disorder are two of the most challenging psychiatric illnesses, affecting millions of people worldwide. 
Schizophrenia is a severe mental disorder characterized by a wide range of symptoms, including delusions, hallucinations, disorganized thinking, and abnormal behaviors \cite{mccutcheon2020schizophrenia}. 
On the other hand, bipolar disorder is a mood disorder characterized by recurrent episodes of mania and depression \cite{grande2016bipolar}. 
Schizophrenia and bipolar disorder are chronic illnesses that can severely impact an individual's daily life and functioning \cite{green2006cognitive}. 
The symptoms of these disorders can be distressing and debilitating, making it difficult for patients to maintain relationships, work, or engage in everyday activities. 
Unfortunately, accurate diagnosis of these disorders is often delayed or missed, resulting in inappropriate or ineffective treatment. 

While the two disorders have distinct clinical features, they also share some similarities in terms of symptoms and genetic risk factors. 
Both disorders register problems in cognitive achievements reporting deficits in visuospatial performance as a precursor of both disorders. 
This overlap has led some researchers to suggest that the two disorders may be part of a broader spectrum of mental illnesses that share underlying genetic and environmental risk factors \cite{maier2006schizophrenia}.

It is essential to distinguish between schizophrenia and bipolar disorder because although they share some common symptoms, they require different treatments. 
Misdiagnosis or delayed diagnosis can lead to inappropriate or ineffective treatments, resulting in poor outcomes for patients. 
For example, antipsychotic medications, which are typically used to treat schizophrenia, may exacerbate symptoms of mania in bipolar disorder \cite{narasimhan2007review}. 
Conversely, mood stabilizers and antidepressants, typically used to treat bipolar disorder, may not be effective for treating symptoms of schizophrenia \cite{nayak2021mood}.

In recent years, the development of new techniques for brain imaging has led to significant advances in the diagnosis and treatment of schizophrenia and bipolar disorder. 
Resting-state functional magnetic resonance imaging (rsfMRI) has emerged as a promising tool for understanding the underlying neural mechanisms of these disorders. 
RsfMRI measures brain activity by detecting changes in blood flow to different regions of the brain during periods of rest. 
Studies have shown that there are distinct patterns of brain activity associated with schizophrenia and bipolar disorder, and these patterns can be used to differentiate between the two disorders \cite{reavis2017assessing}, \cite{rashid2014dynamic}.
In this paper, we present methods for differentiating schizophrenia and bipolar disorder based on rsfMRI data. We apply machine learning algorithms to analyze patterns of resting-state brain activity and evaluate the models' performances on a large dataset of patients with schizophrenia and bipolar disorder. The proposed system was submitted to the IEEE Signal Processing Cup (SPC) 2023. \cite{{psychosis-classification-with-rsfmri}}.
\section{Dataset}
The dataset used in this work was provided by the Brain Space Initiative for the IEEE SPC \cite{psychosis-classification-with-rsfmri}.
It consists of features extracted from the rsfMRI data of individuals with Schizophrenia and Bipolar disorder. 
The dataset was obtained by using 105 intrinsic connectivity network (ICN) time courses derived from a multi-spatial-scale spatially constrained ICA approach and their functional network connectivity (FNC). 

The provided features were extracted using the following steps \cite{psychosis-classification-with-rsfmri}:
\begin{enumerate}
    \item Quality control was applied to identify high-quality data.
    \item Each subject's rsfMRI data were preprocessed using a common procedure, including rigid body motion correction, slice timing correction, and distortion correction. 
    \item Preprocessed subject data were registered into a common space, resampled to~3 mm$^3$ isotropic voxels and spatially smoothed using a Gaussian kernel with a 6~mm full width at half-maximum (FWHM).
    \item A multi-spatial-scale template of 105 ICNs obtained from 100k+ subjects was used and a constrained ICA approach to obtain subject-specific ICN time courses. 
    \item To calculate FNC, ICN time courses were cleaned using a common standard and FNC is estimated by calculating the Pearson correlation between each pair of ICN time courses resulting in one FNC matrix for each individual.  
\end{enumerate}

The training dataset consists of ICN and FCN features for 471 individuals and the test set contains the features of 315 individuals. 
An additional test set was withheld to evaluate the submitted models in the IEEE SPC.

\section{Methods}

We applied different methods for the differentiation of the two diagnostic groups for the FCN and ICN features according to their nature. 
For the differentiation of the FCN features between diagnostic groups, we applied statistical methods for feature selection and machine learning algorithms for binary classifications. 
For the classification of ICNs, we applied digital signal processing techniques as well as machine learning techniques for feature extraction and  binary classification. 
With each of the trained models, we predicted the labels of the test set using soft probability scores, which we used to evaluate the models' AUC scores for the IEEE SPC.  

\subsection{Intrinsic Connectivity Network}
Intrinsic connectivity networks are a set of brain networks, defined based on the intrinsic functional connectivity of different brain regions that are identified using fMRI \cite{laird2011behavioral}. 
They are thought to reflect the underlying organization of the brain which is helpful for understanding brain function \cite{dovern2012intrinsic}. 
ICNs are mainly identified using techniques such as independent component analysis (ICA), which separates the fMRI data into independent components that correspond to different functional networks \cite{zuo2010reliable}, \cite{seeley2007dissociable}. 
Research shows alterations in ICNs in various neuropsychiatric disorders, emphasizing their importance for understanding the neural mechanisms underlying those conditions \cite{mohan2016focus}. 
ICNs provide a powerful tool for investigating the functional organization of the brain and its implications for cognition and behavior \cite{wang2017altered},  which are altered processes in patients with schizophrenia and bipolar disorder.

\subsubsection{Preprocessing}

To compensate for the varying ICN time course lengths provided by the IEEE SPC organizers, we padded shorter components with zeros to match the maximum ICN length of the dataset. 


\begin{figure}[tbp]
    \centerline{\includegraphics[width=\columnwidth]{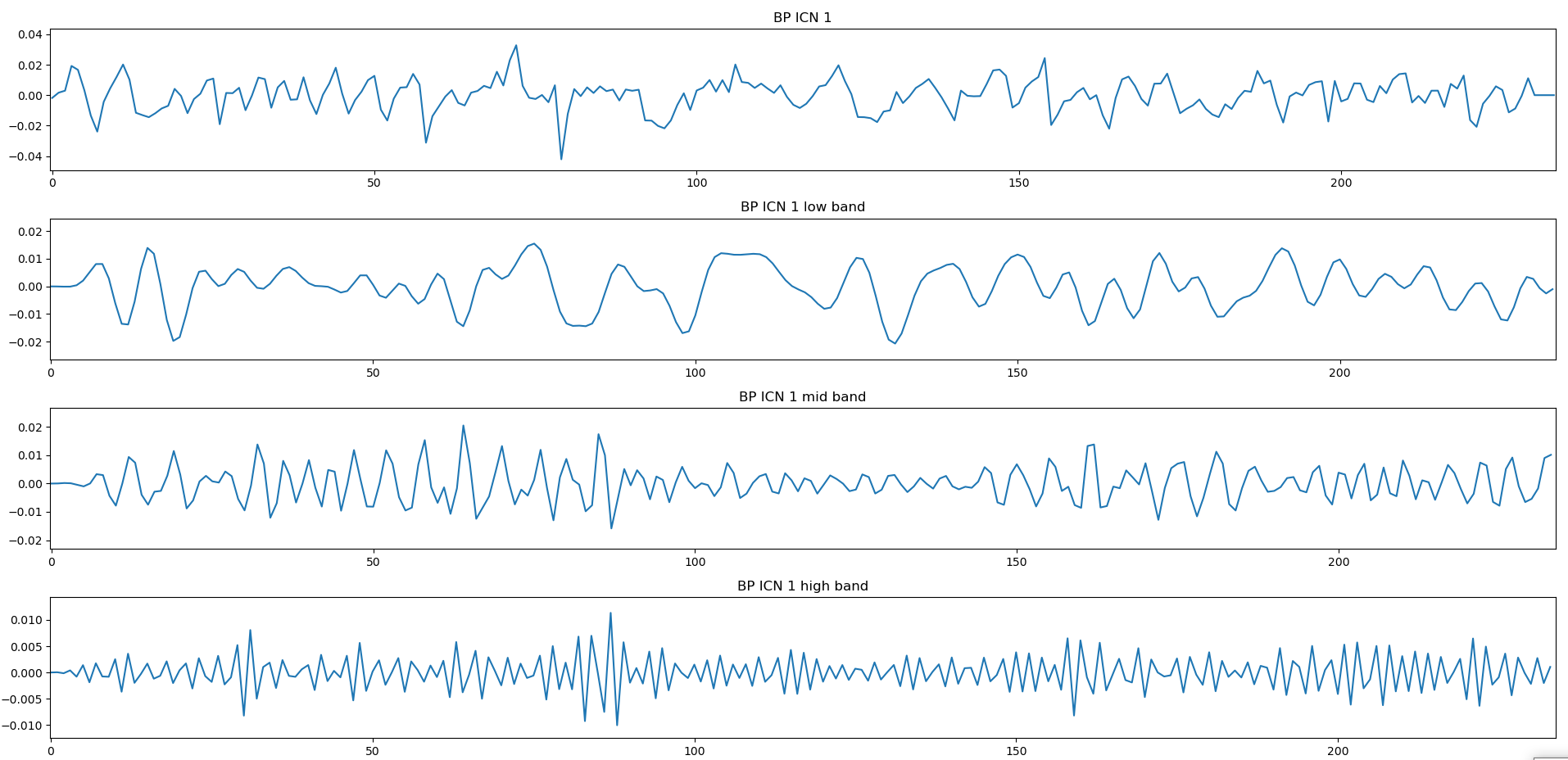}}
    \caption{Sample ICN component of an individual with bipolar disorder (BP) with the original signal and the low, mid, and high band signals obtained with the filter bank}
    \label{fig:icnbp}  
\end{figure}

\begin{figure}[tbp]
    \centerline{\includegraphics[width=\columnwidth]{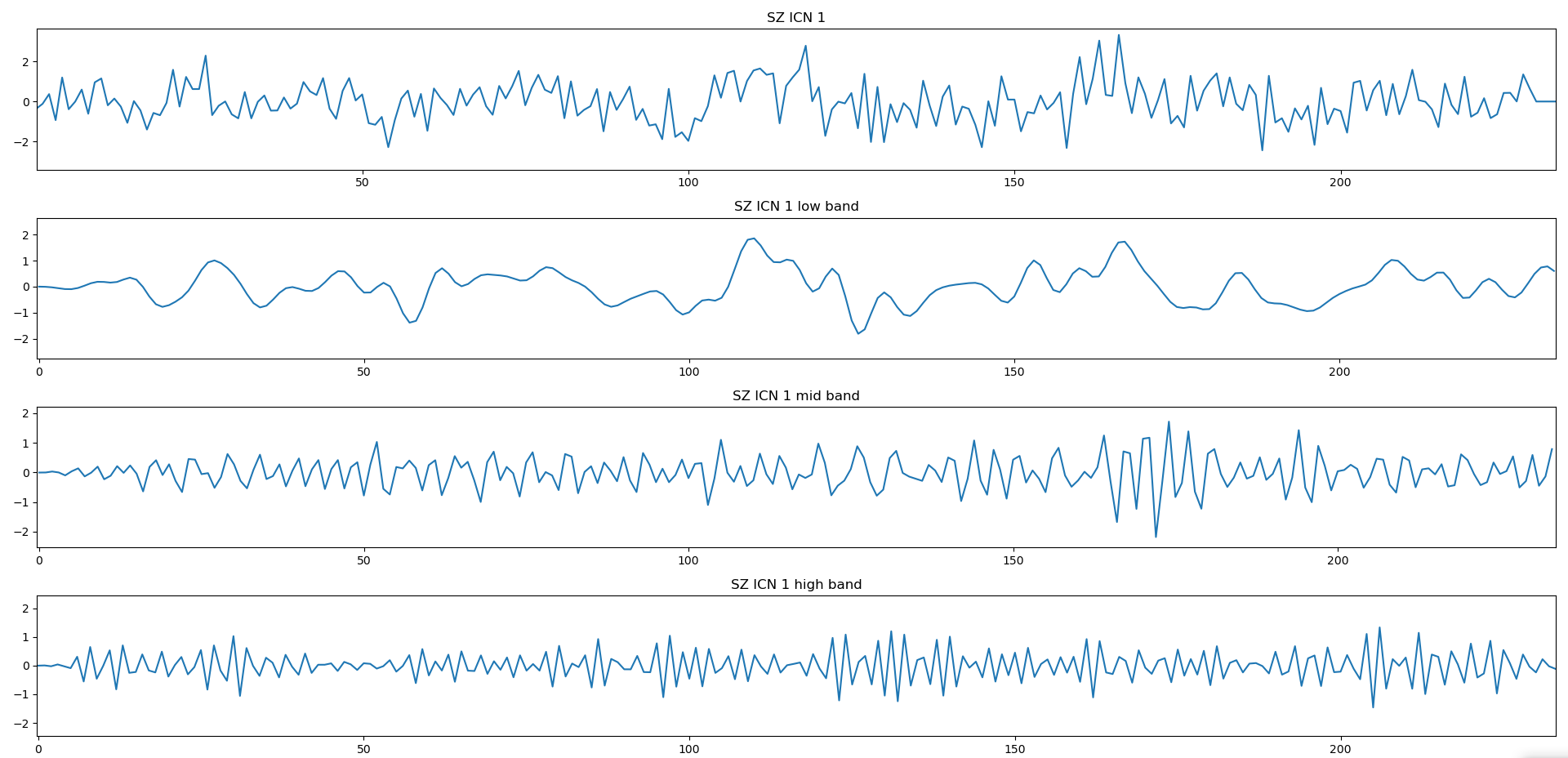}}

    \caption{Sample ICN component of an individual with schizophrenia (SZ), with the original signal and the low, mid, and high band signals obtained with the filterbank }
    \label{fig:icnsz}  
\end{figure}

We used a filter bank to obtain signals in three frequency sub-bands: low, mid, and high. 
The low band contains frequencies between $0.01 - 0.3$~Hz, the mid band contains frequencies between $0.3 - 0.7$~Hz, and the high band frequencies are between $0.7 - 0.99$~Hz. 
The filter bank comprises three bandpass Butterworth IIR filters of order 6. 
In Fig.~\ref{fig:icnbp} and Fig.~\ref{fig:icnsz} we show a sample ICN component of an individual with bipolar disorder (BP) and an individual with schizophrenia (SZ) in the filtered frequency sub-bands. 
    
\subsubsection{Spectrogram} 

For obtaining the time-frequency representation of the ICN time courses, we applied the Short Time Fourier Transformation algorithm (STFT) 
using a sliding Tukey window of length 22. 

We then calculate the power spectrogram. 

By stacking the spectrograms of each ICN, we create a volumetric representation for the individual with dimensions $12 \times 11 \times 105$ where the $x$-axis represents the frequency components, the $y$-axis the time components, and the $z$-axis the number of ICNs as shown in Fig.~\ref{fig:spectrogram}.

\subsubsection{Scalogram} 
For obtaining a 2D representation of the ICNs with a better time-frequency resolution, we applied the Continuous Wavelet transformation (CWT). 
CWT is a formal tool that provides an overcomplete representation of a signal by ``daughter'' wavelets which are scaled and translated copies of the finite-length oscillating waveforms known as the ``mother wavelet''. 

The wavelet analysis provides not only accurate frequency information but at the same time it provides information for accurate time localization of the frequency components. 
This property makes the wavelet transformation highly applicable for the analysis of signals which are characterized by the occurrence of transient events.

\begin{figure}[tbp]
    \centerline{\includegraphics[width=\columnwidth]{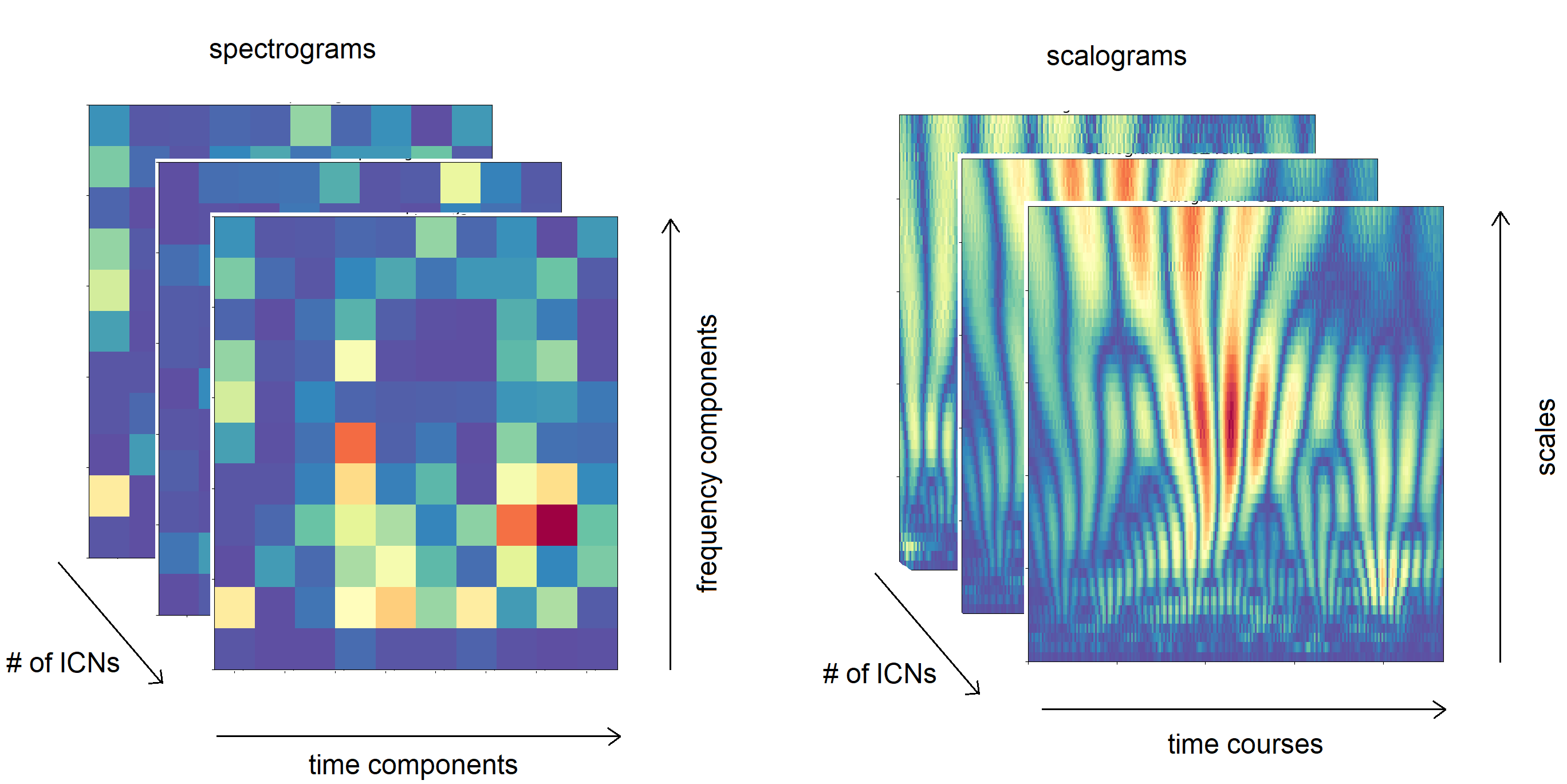}}
    \caption{3D representation of spectrograms (left) and scalograms (right).}
    \label{fig:spectrogram}  
\end{figure}

We generated scalograms of the ICN components using the Morlet wavelet and 50 scales. 
We stacked the scalograms to create 3D information with dimensions $49\times234\times105$ where the $x$-axis represents the scales, the $y$-axis represents the time courses and the $z$-axis represents the number of ICNs as shown in Fig~\ref{fig:spectrogram}.

\begin{figure}[tbp]
    \centerline{\includegraphics[width=\columnwidth]{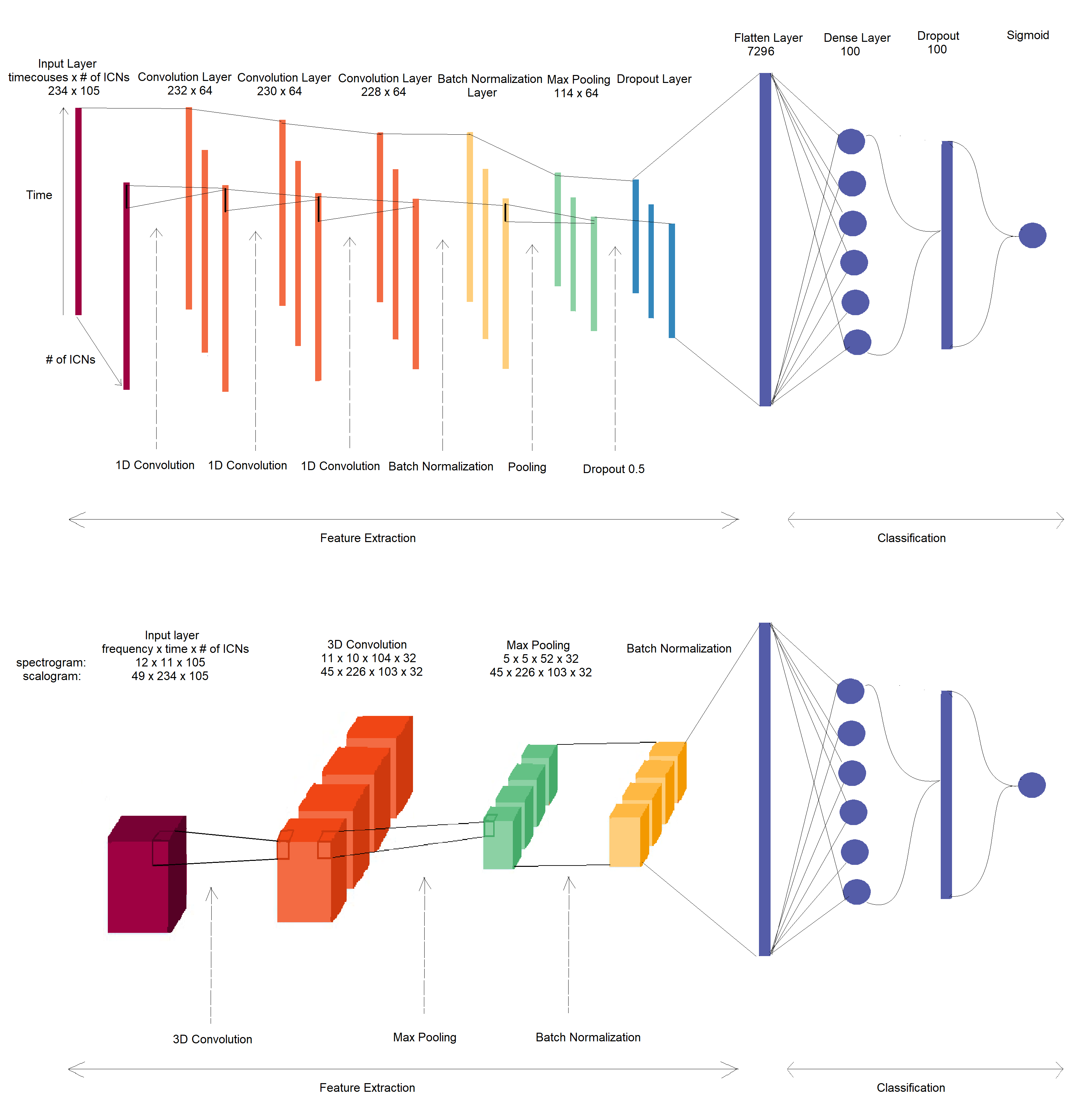}}
    \caption{1D (top) and 3D (bottom) CNN architectures.}
    \label{fig:cnns}
\end{figure}

\subsubsection{Classification}

For the classification of raw ICNs, and filtered ICNs in different bandwidths, we used 1D Convolutional Neural Networks (CNNs). 
The architecture of the model is shown in Fig~\ref{fig:cnns}. 
The outputs of each convolution layer are passed through the ReLU activation function \cite{agarap2018deep}. 

For the classification of the 3D stack of scalograms and spectrograms, we used the 3D CNN model shown in Fig.~\ref{fig:cnns}.
We used a simpler architecture, because of memory and computational constraints. 

We trained both models using Adam \cite{kingma2014adam} in 100 epochs with a batch size of 32. 
To prevent overfitting, we used Early Stopping with patience of 20 Epochs based on the binary cross entropy loss function \cite{prechelt2002early}. 

To evaluate the performance of our models we used the AUC score \cite{ling2003auc} on 20\% of the training dataset.

\subsection{Functional Connectivity Network}

Functional connectivity networks refer to patterns of synchronous activity between different brain regions that are thought to underlie specific cognitive functions \cite{buckner2013opportunities}. 
These networks are typically identified using functional magnetic resonance imaging (fMRI) and other neuroimaging techniques that can measure the correlation between the activity of different brain regions. 
Resting-state networks are commonly studied in the absence of any explicit task or stimulus. 
The study of functional connectivity networks has important implications for understanding the neural basis of complex cognitive processes and for identifying biomarkers of neurological and psychiatric disorders \cite{jafri2008method}. 

\begin{figure}[tbp]
    \centerline{\includegraphics[width=\columnwidth]{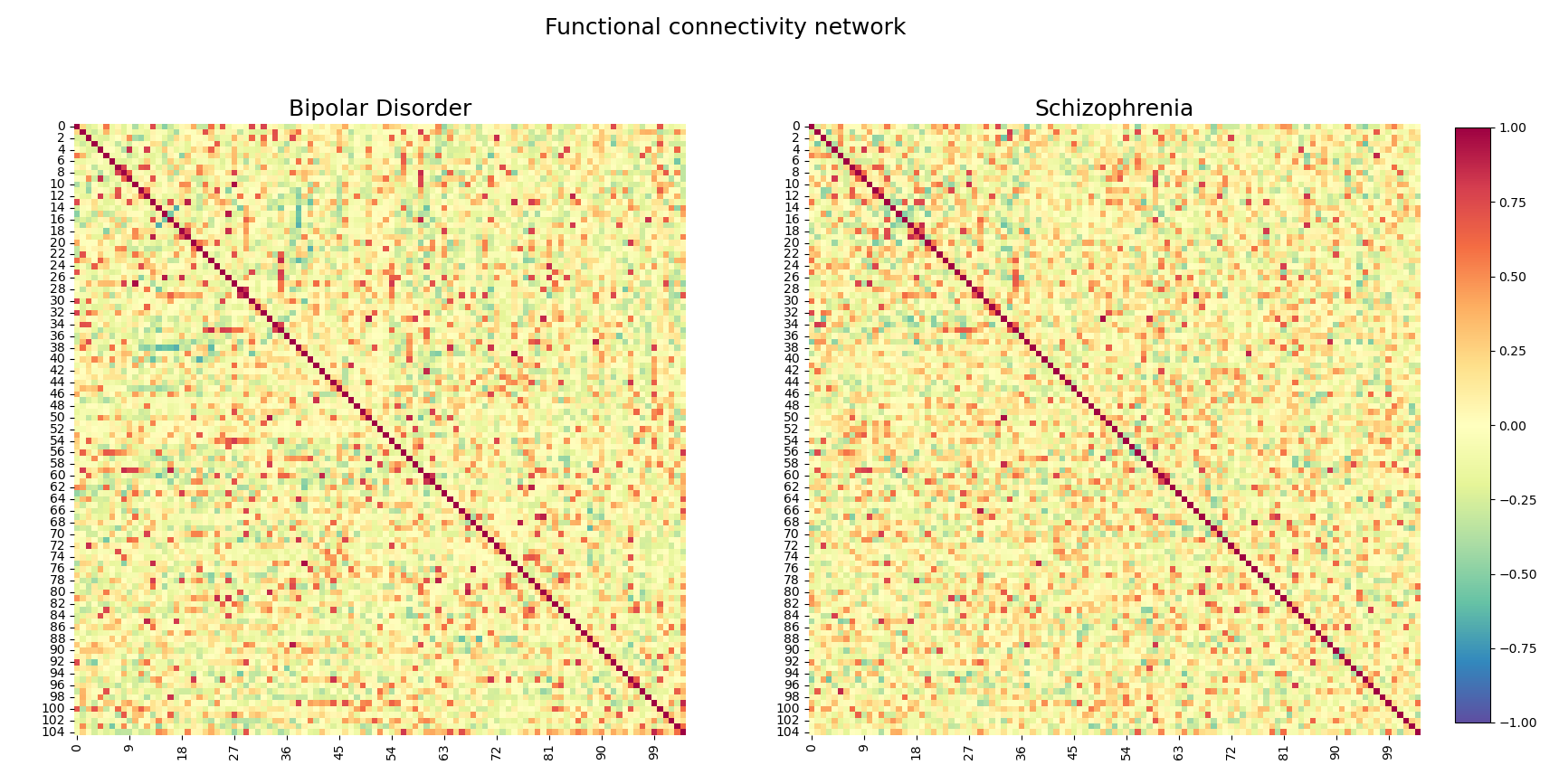}}
    \caption{Functional connectivity network }
    \label{fig:fcn}
\end{figure}

The IEEE SPC Dataset FCNs are generated by calculating Pearson correlation coefficients of the ICN time courses which results in a symmetrical matrix as shown in Fig.~\ref{fig:fcn}. 
The lower triangular matrix is flattened  and
provided in the IEEE SPC dataset in the form of a vector with dimensions $1 \times 5460$. 

\subsubsection{Feature selection} 

We normalized the vector values in the range of 0 to 1 and applied the chi-square test \cite{mchugh2013chi}, for feature selection. 
We selected the 20 best features that explain most of the variability in the dataset.  

\subsubsection{Classification}

We applied different machine-learning algorithms for binary classification of the raw ICN features in the train set, including: Logistic Regression (LR), Support Vector Machines (SVM), Linear Discriminant Analysis (LDA), Gaussian Naive Bayes (GNB), K-Nearest Neighbours (KNN).
For hyperparameters tuning, we did a Grid Search with 5-fold cross-validation \cite{lavalle2004relationship}. 

We applied the same methodology for training the machine learning algorithms using only the selected features.

\subsection{Evaluation}
For the evaluation of the models, the AUC score was used in the IEEE SPC.
We used the best models to predict the soft probability score of the labels of the test set provided in the competition dataset.  The soft probability score indicates the confidence the model has in its prediction and it is used to calculate the AUC score.
For the competition, the AUC was averaged for the public test set, i.e. the Public AUC score, and the withheld test set, i.e. the private AUC Score.

\section{Results}
The AUC scores from our CNN models for the filtered raw ICNs, and stacked ICN spectrograms and scalograms are shown in Table~\ref{tab:icn}.
We can see that the best score is obtained by 1D CNN model trained on raw ICN timecourses. The filtered signal in the range of 0.3 - 0.7 Hz achieved a better AUC score compared to the other bandwidths. Classification of scalograms produced better AUC score, compared to the classification of spectrograms. However, the extracted features from the ICN components achieved slightly worse AUC scores, overall. 
The AUC scores from the classical machine learning models for all of the FCNs and the selected FCNs are shown in Tables~\ref{tab:fcnall} \& \ref{tab:fcnsel}.
We can see that the best score is obtained by the Linear Discriminant Analysis algorithm. Classification of all features resulted in a slightly better overall performance compared to using only the top 20 features. However, while including all features provided enhancement in the classification results,  using a smaller subset of the most relevant features can yield comparable outcomes.

\begin{table}[tbp]
    \caption{Classification of ICN features}
    \begin{center}
        \begin{tabular}{cccc}
            \textbf{Features} & \textbf{Method}& \textbf{AUC Public} & \textbf{AUC Private} \\
            \cline{1-4}
            \hline
            \textbf{raw ICN} & \textbf{1D CNN} & \textbf{0.705} & \textbf{0.637} \\
            ICN $0.01-0.3$ Hz & 1D CNN & 0.509 & 0.551\\
            ICN $0.3-0.7$ Hz & 1D CNN & 0.5881 & 0.597\\
            ICN $0.7-0.99$ Hz & 1D CNN & 0.588 & 0.563\\
            spectrograms  & 3D CNN &  0.509 & 0.551\\
            scalograms & 3D CNN &  0.626 & 0.528\\
        \end{tabular}
        \label{tab:icn}
    \end{center}
\end{table}

\begin{table}[tbp]
    \caption{Classification of all FCN features}
    \begin{center}
        \begin{tabular}{ccc}
            \textbf{Algorithm} & \textbf{AUC Public} & \textbf{AUC Private} \\
            \cline{1-3}
            \hline
            LR & 0.678 & 0.589\\
            SVM &  0.675 & 0.510\\
            \textbf{LDA} &  \textbf{0.686} & \textbf{0.589}\\
            GNB &  0.624 & 0.547\\
            KNN &  0.549 & 0.502\\
            DT &  0.500 & 0.500\\
            RF &  0.658 & 0.540\\
        \end{tabular}
        \label{tab:fcnall}
    \end{center}
\end{table}

\begin{table}[tbp]
    \caption{Classification of selected 20 FCN features}
    \begin{center}
        \begin{tabular}{ccc}
            \textbf{Algorithm} & \textbf{AUC Public} & \textbf{AUC Private} \\
            \cline{1-3}
            \hline
            LR &  0.653 & 0.530 \\
            SVM &  0.634 & 0.520\\
            \textbf{LDA} &  \textbf{0.658} & \textbf{0.525}\\
            GNB &  0.656 & 0.512\\
            KNN &  0.465 & 0.516\\
            DT &  0.574 & 0.475\\
            RF &  0.636 & 0.589\\
        \end{tabular}
        \label{tab:fcnsel}
    \end{center}
\end{table}

\section{Conclusion}
The differentiation between schizophrenia and bipolar disorder is an important task for accurate diagnosis and proper treatment of patients.
We proposed a number of systems based on rs-fMRI features for achieving this objective. The best-trained 1D CNN model trained on raw ICN timecourses achieved the highest AUC score. The FCNs showed potential but their classification achieved slightly worse AUC scores than raw ICN timecourses. 
Overall, the results are promising but future work needs to be done for improving the differentiation between Schizophrenia and Bipolar disorder. Future work will include combining the ICN and FCN features and applying more advanced classification algorithms such as NN, LSTM, GRU, RNN, and transformers.

\bibliographystyle{IEEEtran}
\bibliography{IEEEabrv,IEEEexample}

\begin{thebibliography}{10}
\providecommand{\url}[1]{#1}
\csname url@samestyle\endcsname
\providecommand{\newblock}{\relax}
\providecommand{\bibinfo}[2]{#2}
\providecommand{\BIBentrySTDinterwordspacing}{\spaceskip=0pt\relax}
\providecommand{\BIBentryALTinterwordstretchfactor}{4}
\providecommand{\BIBentryALTinterwordspacing}{\spaceskip=\fontdimen2\font plus
\BIBentryALTinterwordstretchfactor\fontdimen3\font minus
  \fontdimen4\font\relax}
\providecommand{\BIBforeignlanguage}[2]{{%
\expandafter\ifx\csname l@#1\endcsname\relax
\typeout{** WARNING: IEEEtran.bst: No hyphenation pattern has been}%
\typeout{** loaded for the language `#1'. Using the pattern for}%
\typeout{** the default language instead.}%
\else
\language=\csname l@#1\endcsname
\fi
#2}}
\providecommand{\BIBdecl}{\relax}
\BIBdecl

\bibitem{mccutcheon2020schizophrenia}
R.~A. McCutcheon, T.~R. Marques, and O.~D. Howes, ``Schizophrenia—an
  overview,'' \emph{JAMA psychiatry}, vol.~77, no.~2, pp. 201--210, 2020.

\bibitem{grande2016bipolar}
I.~Grande, M.~Berk, B.~Birmaher, and E.~Vieta, ``Bipolar disorder,'' \emph{The
  Lancet}, vol. 387, no. 10027, pp. 1561--1572, 2016.

\bibitem{green2006cognitive}
M.~F. Green, ``Cognitive impairment and functional outcome in schizophrenia and
  bipolar disorder,'' \emph{Journal of Clinical Psychiatry}, vol.~67, p.~3,
  2006.

\bibitem{maier2006schizophrenia}
W.~Maier, A.~Zobel, and M.~Wagner, ``Schizophrenia and bipolar disorder:
  differences and overlaps,'' \emph{Current Opinion in Psychiatry}, vol.~19,
  no.~2, pp. 165--170, 2006.

\bibitem{narasimhan2007review}
M.~Narasimhan, T.~O. Bruce, and P.~Masand, ``Review of olanzapine in the
  management of bipolar disorders,'' \emph{Neuropsychiatric disease and
  treatment}, vol.~3, no.~5, pp. 579--587, 2007.

\bibitem{nayak2021mood}
R.~Nayak, I.~Rosh, I.~Kustanovich, and S.~Stern, ``Mood stabilizers in
  psychiatric disorders and mechanisms learnt from in vitro model systems,''
  \emph{International Journal of Molecular Sciences}, vol.~22, no.~17, p. 9315,
  2021.

\bibitem{reavis2017assessing}
E.~A. Reavis, J.~Lee, J.~K. Wynn, S.~A. Engel, M.~S. Cohen, K.~H. Nuechterlein,
  D.~C. Glahn, L.~L. Altshuler, and M.~F. Green, ``Assessing neural tuning for
  object perception in schizophrenia and bipolar disorder with multivariate
  pattern analysis of fmri data,'' \emph{Neuroimage: Clinical}, vol.~16, pp.
  491--497, 2017.

\bibitem{rashid2014dynamic}
B.~Rashid, E.~Damaraju, G.~D. Pearlson, and V.~D. Calhoun, ``Dynamic
  connectivity states estimated from resting fmri identify differences among
  schizophrenia, bipolar disorder, and healthy control subjects,''
  \emph{Frontiers in human neuroscience}, vol.~8, p. 897, 2014.

\bibitem{psychosis-classification-with-rsfmri}
\BIBentryALTinterwordspacing
{Brain Space Initiative}, ``{Psychosis classification with rsfMRI},'' {2023}.
  [Online]. Available:
  \url{{https://kaggle.com/competitions/psychosis-classification-with-rsfmri}}
\BIBentrySTDinterwordspacing

\bibitem{laird2011behavioral}
A.~R. Laird, P.~M. Fox, S.~B. Eickhoff, J.~A. Turner, K.~L. Ray, D.~R. McKay,
  D.~C. Glahn, C.~F. Beckmann, S.~M. Smith, and P.~T. Fox, ``Behavioral
  interpretations of intrinsic connectivity networks,'' \emph{Journal of
  cognitive neuroscience}, vol.~23, no.~12, pp. 4022--4037, 2011.

\bibitem{dovern2012intrinsic}
A.~Dovern, G.~R. Fink, A.~C.~B. Fromme, A.~M. Wohlschl{\"a}ger, P.~H. Weiss,
  and V.~Riedl, ``Intrinsic network connectivity reflects consistency of
  synesthetic experiences,'' \emph{Journal of Neuroscience}, vol.~32, no.~22,
  pp. 7614--7621, 2012.

\bibitem{zuo2010reliable}
X.-N. Zuo, C.~Kelly, J.~S. Adelstein, D.~F. Klein, F.~X. Castellanos, and M.~P.
  Milham, ``Reliable intrinsic connectivity networks: test--retest evaluation
  using ica and dual regression approach,'' \emph{Neuroimage}, vol.~49, no.~3,
  pp. 2163--2177, 2010.

\bibitem{seeley2007dissociable}
W.~W. Seeley, V.~Menon, A.~F. Schatzberg, J.~Keller, G.~H. Glover, H.~Kenna,
  A.~L. Reiss, and M.~D. Greicius, ``Dissociable intrinsic connectivity
  networks for salience processing and executive control,'' \emph{Journal of
  Neuroscience}, vol.~27, no.~9, pp. 2349--2356, 2007.

\bibitem{mohan2016focus}
A.~Mohan, A.~J. Roberto, A.~Mohan, A.~Lorenzo, K.~Jones, M.~J. Carney,
  L.~Liogier-Weyback, S.~Hwang, and K.~A. Lapidus, ``Focus: the aging brain:
  the significance of the default mode network (dmn) in neurological and
  neuropsychiatric disorders: a review,'' \emph{The Yale journal of biology and
  medicine}, vol.~89, no.~1, p.~49, 2016.

\bibitem{wang2017altered}
L.~Wang, H.~Shen, Y.~Lei, L.-L. Zeng, F.~Cao, L.~Su, Z.~Yang, S.~Yao, and
  D.~Hu, ``Altered default mode, fronto-parietal and salience networks in
  adolescents with internet addiction,'' \emph{Addictive Behaviors}, vol.~70,
  pp. 1--6, 2017.

\bibitem{agarap2018deep}
A.~F. Agarap, ``Deep learning using rectified linear units (relu),''
  \emph{arXiv preprint arXiv:1803.08375}, 2018.

\bibitem{kingma2014adam}
D.~P. Kingma and J.~Ba, ``Adam: A method for stochastic optimization,''
  \emph{arXiv preprint arXiv:1412.6980}, 2014.

\bibitem{prechelt2002early}
L.~Prechelt, ``Early stopping-but when?'' in \emph{Neural Networks: Tricks of
  the trade}.\hskip 1em plus 0.5em minus 0.4em\relax Springer, 2002, pp.
  55--69.

\bibitem{ling2003auc}
C.~X. Ling, J.~Huang, H.~Zhang \emph{et~al.}, ``Auc: a statistically consistent
  and more discriminating measure than accuracy,'' in \emph{Ijcai}, vol.~3,
  2003, pp. 519--524.

\bibitem{buckner2013opportunities}
R.~L. Buckner, F.~M. Krienen, and B.~T. Yeo, ``Opportunities and limitations of
  intrinsic functional connectivity mri,'' \emph{Nature neuroscience}, vol.~16,
  no.~7, pp. 832--837, 2013.

\bibitem{jafri2008method}
M.~J. Jafri, G.~D. Pearlson, M.~Stevens, and V.~D. Calhoun, ``A method for
  functional network connectivity among spatially independent resting-state
  components in schizophrenia,'' \emph{Neuroimage}, vol.~39, no.~4, pp.
  1666--1681, 2008.

\bibitem{mchugh2013chi}
M.~L. McHugh, ``The chi-square test of independence,'' \emph{Biochemia medica},
  vol.~23, no.~2, pp. 143--149, 2013.

\bibitem{lavalle2004relationship}
S.~M. LaValle, M.~S. Branicky, and S.~R. Lindemann, ``On the relationship
  between classical grid search and probabilistic roadmaps,'' \emph{The
  International Journal of Robotics Research}, vol.~23, no. 7-8, pp. 673--692,
  2004.

\end{thebibliography}

\end{document}